# Generalized Huang's Equation for Phonon Polariton in Polyatomic Polar Crystal

Weiliang Wang[1,4,5], Ningsheng Xu[2,3,4], Yingyi Jiang[2,3,4], Zhibing Li[3,4,6], Zebo Zheng[2,3,4], Huanjun Chen[2,3,4*], and Shaozhi Deng[2,3,4*]

[1]School of Physics, Sun Yat-sen University, Guangzhou 510275, China
[2]School of Electronics and Information Technology, Sun Yat-sen University, Guangzhou 510275, China
[3]State Key Laboratory of Optoelectronic Materials and Technologies, Sun Yat-sen University, Guangzhou 510275, China
[4]Guangdong Province Key Laboratory of Display Material and Technology, Sun Yat-sen University, Guangzhou 510275, China
[5]Center for Neutron Science and Technology, Sun Yat-sen University, Guangzhou 510275, China
[6]School of Science, Shenzhen Campus, Sun Yat-sen University, Shenzhen 518107, China
Email: chenhj8@mail.sysu.edu.cn (H. Chen); stsdsz@mail.sysu.edu.cn (S. Deng)

**Abstract:** The original theory of phonon polariton is Huang's equation which is suitable for diatomic polar crystals only. We proposed a generalized Huang's equation without fitting parameters for phonon polariton in polyatomic polar crystals. We obtained the dispersions of phonon polariton in GaP (bulk), hBN (bulk and 2D), α-$MoO_3$ (bulk and 2D) and $ZnTeMoO_6$ (2D), which agree with the experimental results in the literature and of ourselves. We also obtained the eigenstates of the phonon polariton. We found that the circular polarization of the ion vibration component of these eigenstates is nonzero in hBN flakes. The result is different from that of the phonon in hBN.

TOC Graphic

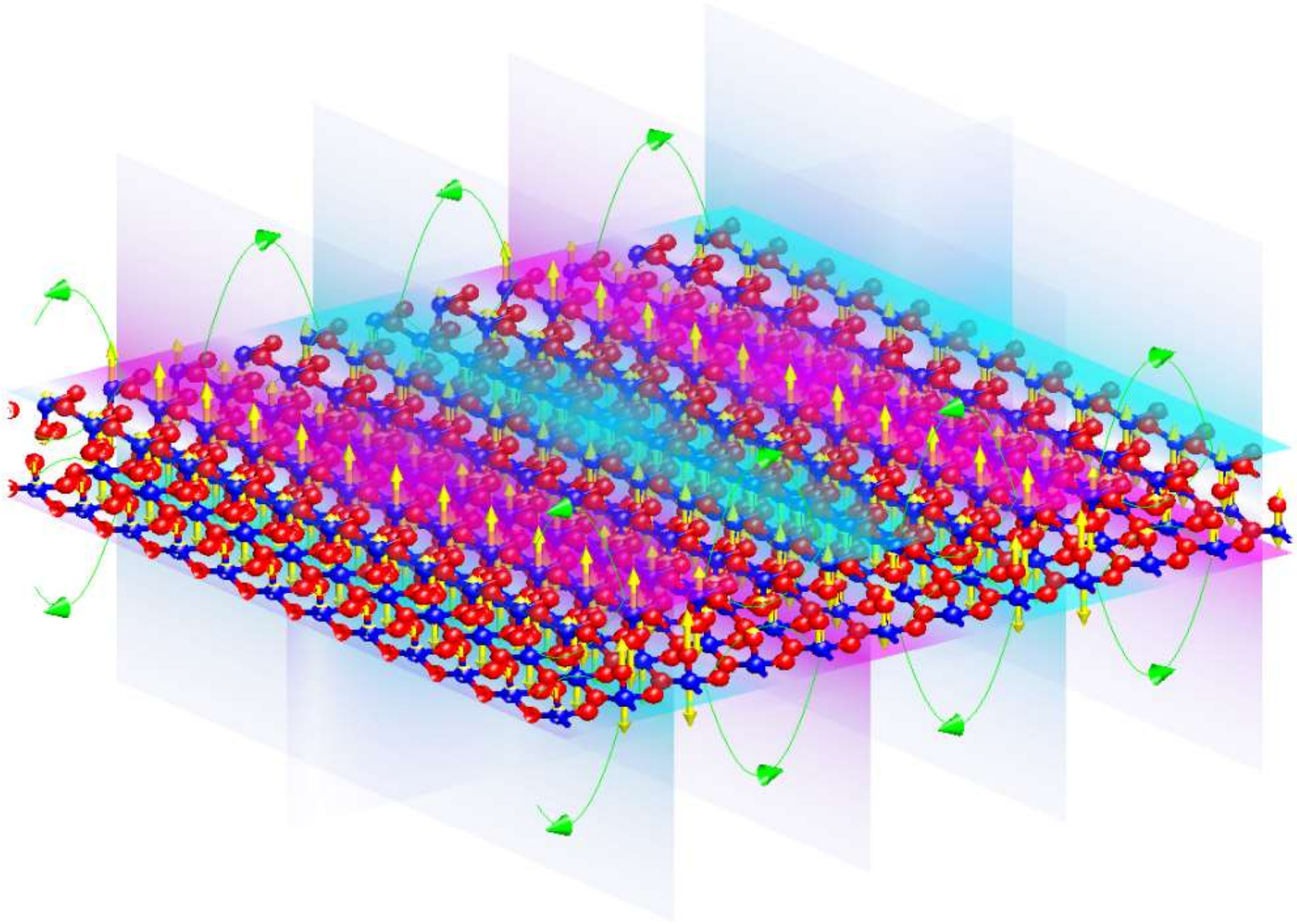

Schematic illustration of PhP propagating along a 2D crystal. The yellow arrows indicate the vibration of atoms. The green lines with arrows indicate the electric displacement vector. The color map displays the component of the electric displacement vector in the direction perpendicular to the 2D crystal plane.

## I. Introduction

Phonon polaritons (PhPs) are new vibration modes formed by the coupling between the lattice vibration (optical phonon) and electromagnetic waves in polar crystals [1]. They are important primary excitations in polar crystals. In recent years, PhPs have attracted more and more attention due to the development of nanophotonic technology, especially in low-dimensional materials and structures. They have important applications in mid-infrared photodetection[2], coherent thermal light source[3], enhancing IR light-matter interaction[4, 5], high-density IR data storage[6], sencing[7-9], enhancing near field radiative heat transfer[10], radiative cooling[11], heat dissipation[12], developing metamaterials[13], frequency tunable terahertz wave generation.[14, 15], topological photonics[16-18], and enhancing dynamical vacuum effects[19].

The first reason why PhPs have such important application prospects is that their dominant loss mechanism is the scattering between optical phonons. Therefore, their average lifetime can reach more than one picosecond, which is far longer than that of other polaritons (such as plasmon polaritons).[20] Secondly, PhP characteristics can be tuned by tailing the material (e. g. stacking and nanostructuring) [21] and by intercalation[22, 23]. Thirdly, PhP materials have special dielectric functions. Specifically, in some specific frequency ranges, its dielectric function is epsilon-near-zero (ENZ), which can be used in replacing metamaterials with complicated structures;[24] and it can behave as a hyperbolic medium (one of the principal components of their dielectric tensor has the opposite sign to the other two principal components), whose isofrequency surface in wave-vector space is hyperbolic, then the PhP is called hyperbolic phonon polariton (HPhP). [25] HPhP can result in peculiar optical phenomena, e. g. high momentum[26], low group velocity[27, 28], negative phase velocity[29], ultra-long lifetime [30], highly directional propagation[31-33], and concave (anomalous) wavefronts[34]. These can lead to sub-wavelength electromagnetic field localization[35] and negative refraction[36, 37], which has essential application potential in nanophotonics devices.[21, 38-40]

There were three main theoretical methods to describe the properties of PhPs. The first one is the famous Huang's equation [1], which is suitable for diatomic polar crystals (whose unit cell consists of one positive ion and one negative ion) only and not for polyatomic polar crystals. The second one is interpreting experimental results with phenomenological models. [31, 41-43]. To ensure their applicability, phenomenological theories usually contain a large number of fitting parameters, which makes them difficult to use. It cannot reveal the deep physical meaning behind the phenomenon to clarify the mechanism of crystal structure influence on phonon polaritons. The third one is calculating the dielectric function with the Lorentz model [22, 32, 44-48], then calculating the electromagnetic wave propagation in polar crystals with Maxwell's equation. This method did not consider the effect of electromagnetic field variation in space on dielectric function, so it did not fully consider the electromagnetic wave. The wave vector direction dielectric tensor partly solved this problem[49]. However, this theory can not get the eigenstates of PhPs, so it can not give us the relationship between the ion vibration mode and the polarization state and frequency of the electromagnetic wave, so it is difficult to investigate the relationship between the properties of PhPs and the crystal structure. Therefore, developing a theory to fully describe PhPs in polyatomic polar crystals without fitting parameters is demanding.

In recent years, PhPs in two-dimensional atomic crystals (e. g. hexagonal boron nitride and α-$MoO_3$) were found to have plenty of peculiar characteristics, such as high in-plane localization, various tuning ways (electrical, optical, thermodynamic). It brings new opportunities for enhancing light-matter interactions and tunable optoelectronic devices at the nanoscale.[22, 50-53] Therefore, a theory suitable for two-dimensional crystals besides bulk crystals is demanding. The present paper generalized Huang's equation for this goal.

## II. Generalized Huang's equation

The original theory of PhP is Huang's equation[1]:

$$\ddot{\vec{W}} = b_{11}\vec{W} + b_{12}\vec{E} \tag{1}$$

$$\vec{P} = b_{21}\vec{W} + b_{22}\vec{E} \tag{2}$$

where $\vec{W}$ is the relative displacement between positive ion and negative ion, $\vec{P}$ is the macroscopic electric polarization, $\vec{E}$ is the macroscopic electric field. The coefficients $b_{11}, b_{12}, b_{21}, b_{22}$ can be determined by the static dielectric constant $\varepsilon(0)$, high-frequency dielectric function $\varepsilon(\infty)$, and the long wave transverse optical phonon frequency $\omega_T$. By combining Huang's equation with Maxwell's electromagnetic field equations, the vibration frequency and eigenstates of lattice vibration coupled with electromagnetic field oscillation can be obtained. This theory was validated by C. H. Henry's Raman experiment on GaP[54]. However, this theory is suitable for diatomic polar crystals only because $\overrightarrow{W}$ is the relative displacement between the positive ion and the negative ion. In polyatomic polar crystals, Eq. (1) and (2) should be replaced with:

$$m_{j'}\Delta\ddot{r}_{\delta}(j'l',t) = -\sum_{\alpha jl}\Phi_{\alpha\delta}(jl,j'l') * \Delta r_{\alpha}(jl,t) + e\sum_{\gamma}Z_{\gamma\delta}(j'l') * E_{\gamma} \tag{3}$$

$$P_{\delta} = \frac{e}{\Omega}\sum_{\mu j'}Z_{\delta\mu}(j'0)\Delta r_{\mu}(j'0,t) + \sum_{\gamma}[\varepsilon_{\infty\delta\gamma} - \delta_{\delta\gamma}]\varepsilon_0 E_{\gamma} \tag{4}$$

where $m_{j'}$ is the mass of the $j'$-th atom, $\Delta r_{\alpha}(jl,t)$ is the displacement of the $j$-th atom in the $l$-th unit cell along direction $\alpha$ at time $t$, $\Phi$ is the second order force constants matrix (second derivatives of potential energy with respect to atomic displacements), $e$ is the electron charge, $Z$ is the Born effective charge tensor[44], $E_{\gamma}$ is the electric field component in $\gamma$ direction, $P_{\beta}$ is the electric polarization in $\beta$ direction, $\Omega$ is the volume of the unit cell, $\varepsilon_{\infty}$ is the dielectric tensor contributed by electrons, $\varepsilon_0$ is the vacuum dielectric constant. $\Phi$, $Z$, and $\varepsilon_{\infty}$ can be obtained with density functional theory calculation (section I, Supporting information). We call Eq. (3) and (4) generalized Huang's equation. Combining the generalized Huang's equation with Maxwell's electromagnetic field equations, the PhPs in polyatomic polar crystals can be appropriately described.

## III. PhPs in bulk materials

Firstly we solve the generalized Huang's equation and Maxwell's equation in bulk materials (section II, Supporting information). GaP is a diatomic polar crystal. The PhP dispersion we obtained for GaP (Figure 1a) agrees well with the experimental results in Ref. [54]. The electric field perpendicular to the wave vector direction is stronger when the PhP frequency is further away from the phonon frequency (Figure S1), which means the PhP is more like a photon when its frequency is farther away from the phonon frequency. The electric field in the longitude optical (LO) phonon is completely parallel to the wave vector direction. There is a pure transverse optical (TO) phonon at 10.9 THz which has no electric field. This pure TO is not shown in the figures.

For polyatomic polar crystals, we obtained the PhP dispersion in bulk hBN (Figures 1b and 1c) and bulk α-$MoO_3$ (Figures 1 d-f). The hBN has four atoms in each unit cell. There are two different PhPs when the wave vector is in-plane (denoted as *x*- direction, shown in Figure 1b). The blue symbols in Figure 1b correspond to out-of-plane atomic vibration (top panel of Figure S2) while the red symbols in Figure 1b correspond to in-plane atomic vibration in the *y* direction (bottom panel of Figure S2). Both of these atomic vibration directions are perpendicular to the wave vector direction. The difference between the two branches of the same symbol color in Figure 1b is the phase difference between the atomic vibration and the electric field oscillation. When the wave vector is out-of-plane (denoted as *z*-direction), there is only one doubly degenerated PhP (red symbols in Figure 1c), in which the atomic vibrations are in-plane (the inset in Figure 1c).

The α-$MoO_3$ has 16 atoms in each unit cell. There are nine PhPs when the wave vector $\vec{q}$ is along [100] (denoted as *x*) direction (Figure 1d) or [010] (denoted as *z*) direction (Figure 1f), respectively. These nine PhPs can be divided into two groups: one group has three PhPs (symbols in red, pink and cyan in Figure 1d and 1f) whose atomic vibration and electric field are along [001] (denoted as *y*) direction (the electric field intensity is shown in Figure S3); the other group has six PhPs (symbols in other colors except red, pink and cyan in Figure 1d and 1f) whose atomic vibration is in the *x*-*z* plane, and the electric field is along *z* (*x*) direction when the wave vector is along *x* (*z*) direction (the electric field intensity is shown in Figure S3). The atomic vibration of the first group PhPs with $\vec{q}$ along *x* and *z* direction are the same (along the *y*-direction, shown in Figure S4). The atomic vibration of the second group PhPs with $\vec{q}$ along *x* and *z* direction are different (shown in Figure S5 and S6, respectively).

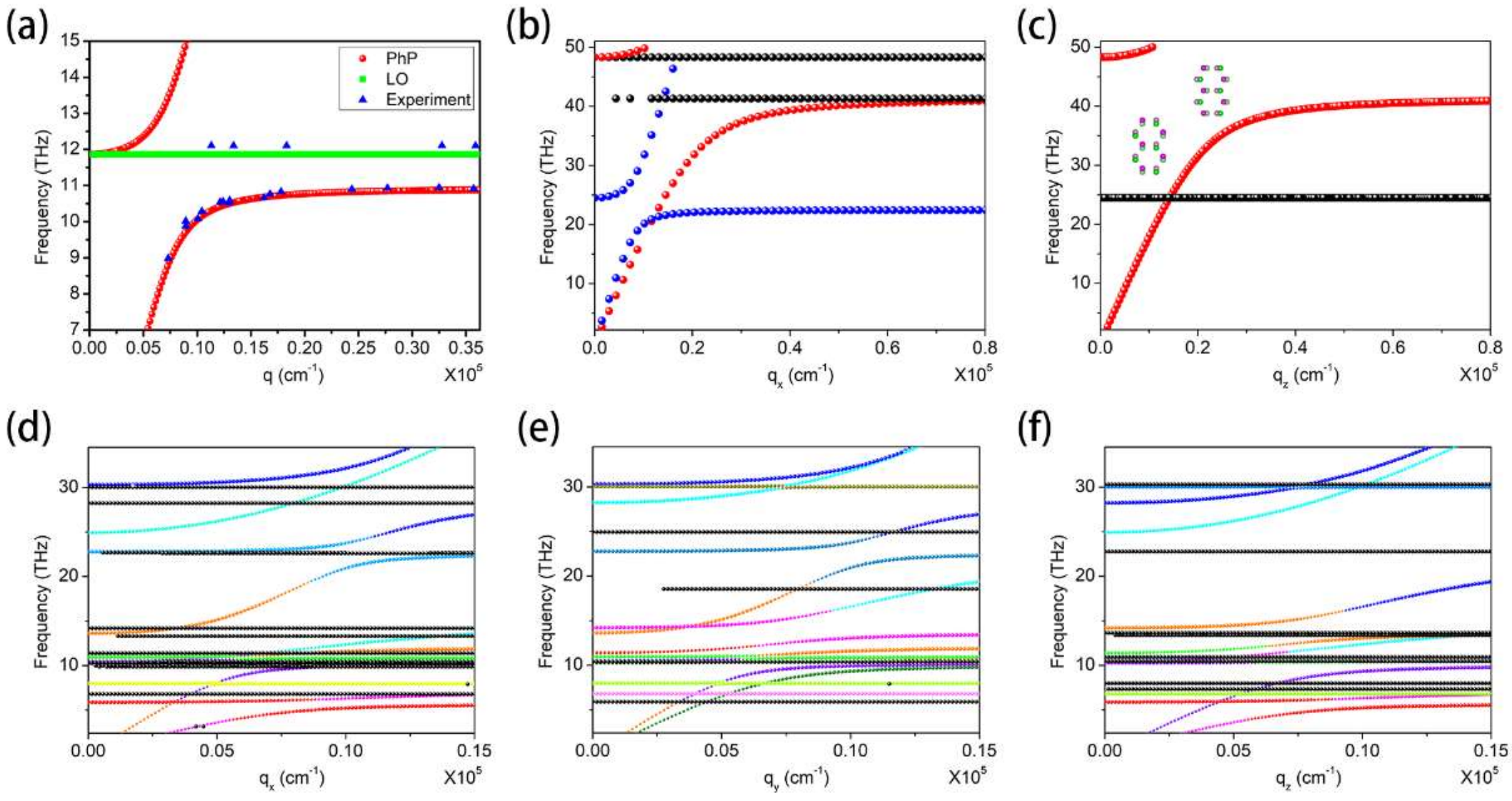

Figure 1. (a) Experimental (blue triangles, from Ref. [54]) and theoretical (red balls) PhP dispersion in GaP, (b) PhP dispersion in bulk hBN when the wave vector direction

is in-plane, (c) PhP dispersion in bulk hBN when the wave vector direction is out-of-plane, the purple (green) balls in the inset refer to boron (nitrogen) atoms, (d, e, f) PhP dispersion in bulk α-$MoO_3$ when the wave vector direction is along [100] (denoted as $x$) direction, [001] (denoted as $y$) direction and [010] (denoted as $z$) direction, respectively. The black horizontal lines are LO phonons. We projected each PhP to phonon modes. The colors of the symbols correspond to phonon mode with max projection. The size of the symbols is proportional to the max projection of the PhPs to the phonon modes.

## IV. PhPs in 2D materials

Hexagonal boron nitride (hBN) and α-$MoO_3$ are the two most widely studied 2D PhP materials, while $ZnTeMoO_6$ is a newly found vdW quarternary oxide that possesses hyperbolic PhPs. Therefore, we calculated their PhP dispersions and compared them with experimental results to validate our generalized Huang's equation (the calculation method is described in section III of the supporting information). Ning Li et al. used monochromatic electron energy-loss spectroscopy in a scanning transmission electron microscope to measure PhP dispersion in hBN flakes with different thicknesses, ranging from monolayer to (~3, 4, and 10 nm) thick samples[27]. Without any fitting parameters, our theoretical results agree with their experimental results (figure 2a, 2b and figure S7). Our theoretical result of the monolayer case performs better than the theoretical results in Ref. [27] and [55], whose slopes are too large. Although the theoretical results of the multilayer case in Ref. [27] also agree well with the experimental results, their theory needs several fitting parameters. Only one or two PhP branches are shown for thin hBN flakes (Figrue 2a and S7), while several PhP branches are shown for the hBN flake with a thickness of 10 nm (Figure 2b). The real part of the wave vector along $z$ (the direction perpendicular to the vdW plane) is close to $n\pi/d$ for the $n^{\text{th}}$ branch from the top in Figure 2b. Hereinafter referred to as $n$ is the index of the branch.

Our theory also gives us the eigenstates, consisting of ion vibration and electromagnetic field oscillation components. Then we can calculate the circular polarization [56, 57] of the ion vibration component of these eigenstates. The calculation method is described in section IV of the supporting information. We found that the circular polarizations along $z$ (denoted as $s_z$) are nonzero in hBN nanoflakes (Figure 2c). The distribution of $s_z$ of PhP in reciprocal space is different from that of phonon. Zhang et al. found that the max $s_z$ of phonons ($\pm\hbar$) in monolayer hexagonal lattices is located at K and K' valleys [56]. The $s_z$ of PhP of hBN nanoflakes near the Γ point is not sensitive to the wave vector (Figures 2a, 2b and S7) but depends on the frequency and the index of the branch.

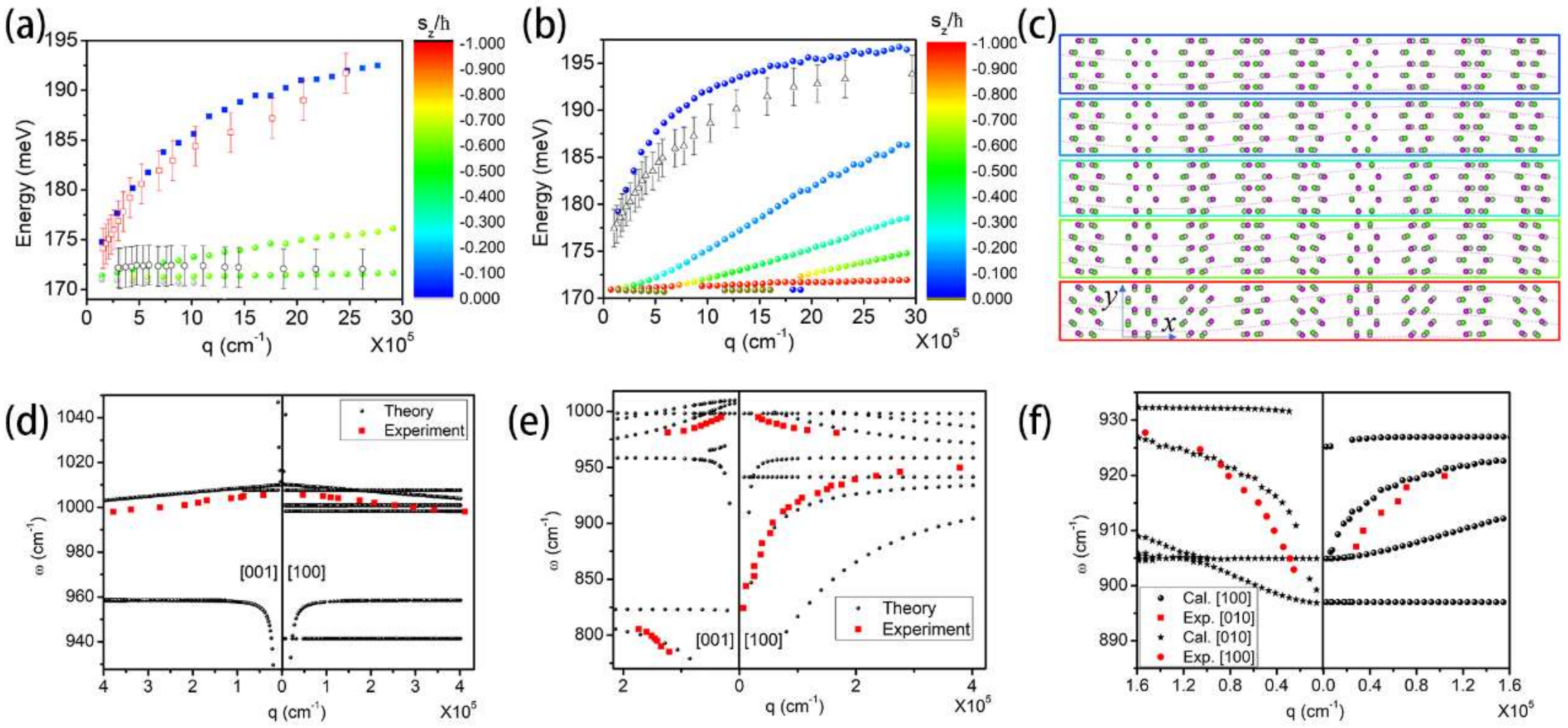

Figure 2. (a) Experimental (black empty circles and red empty squares, from Ref. [27]) and theoretical (balls and solid squares) PhP dispersion in monolayer hBN (circles and balls) and hBN flake with thickness 4 nm (squares). (b) Experimental (black empty triangles, from Ref. [27]) and theoretical (balls) PhP dispersion in hBN flake with thickness 10nm. The color of the balls and the solid squares in (a) and (b) indicates the circular polarization of the ion vibration component of the PhP eigenstates. The wave vectors are in-plane. (c) Atomic vibration in hBN flakes. The colors of the rectangles surrounding the panels correspond to the colors of the symbols in (a) and (b). The purple (green) balls refer to boron (nitrogen) atoms. The wave vectors are in the *x*-direction. Because the wavelength of PhP is much longer than the lattice constants, the ions along the wave vector direction (*x*-direction) are depicted only in every *N* lattice, with *N* being of the order of a thousand or more. (d) Experimental (red squares) and theoretical (black balls) PhP dispersion in α-$MoO_3$ flake with thickness 22 nm. (e) The same as (d), but the thickness is 220 nm. The experimental data in (e) are from Ref. [58]. (f) Experimental (black balls and stars, from Ref. [59]) and theoretical (red squares and solid circles) PhP dispersion in $ZnTeMoO_6$ flake with thickness 190 nm.

To demonstrate the capability of our theory, we utilized the scattering-type scanning near-field optical microscope (s-SNOM) to excite and image very thin α-$MoO_3$ flake (thickness 22 nm) in the mid-infrared regime (Section V, Supporting information). The experimental results agree well with the theoretical results (Figure 2d). The theoretical results also agree with previous experiment results of PhP dispersion in much thicker α-$MoO_3$ flakes with a thickness of 220 nm in Ref.[58] (Figure 2e).

The PhP dispersion of $ZnTeMoO_6$ nanoflake obtained with our theory also agrees well with the experimental results in Ref. [59]. The $ZnTeMoO_6$ crystal comprises 4 elements and 18 atoms in each unit cell. This demonstrates the capability of our method in complicated crystals. The $s_z$ of the PhP eigenstates is zero in α-$MoO_3$ and $ZnTeMoO_6$, which is different from hBN flakes.

## V. Conclusion

We generalized Huang's equation to describe PhP in polyatomic polar crystals. There is no fitting parameter in our theory. All the parameters can be obtained with the first principles calculation. Our theory is validated by comparing our theoretical results with experimental results of GaP (Ref. [54]), hBN (Ref. [27]), α-$MoO_3$ (our experiment in this work and Ref. [58]) and $ZnTeMoO_6$ (Ref. [59]), ranging from diatomic polar crystal to quarternary oxide with 18 atoms in each unit cell, from bulk to 2D material. We also found that the circular polarization $s_z$ of the ion vibration component of the PhP eigenstates is nonzero in hBN flakes. The $s_z$ of PhP in hBN flakes is found to be not sensitive to the wave vector but depends on the frequency and the index of the branch, which is different from phonon.

## Acknowledgments

This work was supported by the National Natural Science Foundation of China (no. 91963205), the Science and Technology Planning Project of Guangdong Province (2023B1212060025), and the Physical Research Platform (PRP) in School of Physics, SYSU.

**Supporting Information**

**Section I.**

**Density functional theory calculation.** The second-order force constants *Φ,* Born effective charge tensor *Z* and dielectric tensor contributed by electrons $\varepsilon_\infty$ are obtained with density functional theory implemented in the Vienna Ab initio simulation package (VASP)[1]. These parameters for 2D structures are obtained from the bulk calculation, while these parameters for monolayer structures are obtained from monolayer structures themself. The electron-core interactions were treated in the projector-augmented wave (PAW) approximation[2]. Grimme's correction scheme (PBE functional including a dispersion correction, PBE-D2) is applied for $MoO_3$. Kinetic energy cutoffs were 520 eV and 500 eV in geometry optimization and other calculations, respectively. All atoms were fully relaxed until the total (free) energy change was smaller than $10^{-8}$ eV. Vacuum slabs of 1.5 nm were inserted between neighboring 2D atom sheets before geometry optimization of monolayer hBN. The second-order force constants and the phonon dispersions were obtained with the Phonopy code.[3] The supercell size, k-point mesh and exchange-correlation potential are listed in table S1.

Table S1. Calculation parameters.

| Material | Exchange-correlation potential | k point mesh | | | Supercell size |
|---|---|---|---|---|---|
| | | geometry optimization | force constant calculation | dielectric tensor calculation | |
| Bulk GaP | CA[4] | 8 × 8 × 8 | 2 × 2 × 2 | 8 × 8 × 8 | 4 × 4 × 4 |
| Monolayer hBN | CA | 15 × 15 × 1 | 3 × 3 × 1 | 15 × 15 × 1 | 5 × 5 × 1 |
| Bulk hBN | CA | 10 × 10 × 4 | 3 × 3 × 3 | 10 × 10 × 4 | 4 × 4 × 1 |
| Bulk $MoO_3$ | PBE[5] | 4 × 4 × 1 | 1 × 1 × 1 | 8 × 8 × 3 | 4 × 4 × 1 |
| $ZnTeMoO_6$ | PBE | 6 × 4 × 4 | 2 × 2 × 2 | 16 × 15 × 9 | 3 × 3 × 2 |

**Section II.**

**Solutions of the generalized Huang's equation for bulk material.**

Maxwell's equations are written as

$$\nabla\times\vec{E} = -\mu_0\frac{\partial\vec{H}}{\partial t} \tag{S1}$$

$$\nabla\times\vec{H} = \frac{\partial}{\partial t}\left(\varepsilon_0\vec{E}+\vec{P}\right) \tag{S2}$$

$$\nabla\cdot\vec{D} = 0 \tag{S3}$$

$$\nabla\cdot\vec{H} = 0 \tag{S4}$$

The solution can be written as

$$P = P_0\exp[i(\vec{q}\cdot\vec{r}-\omega t)] \tag{S5}$$

$$E = E_0\exp[i(\vec{q}\cdot\vec{r}-\omega t)] \tag{S6}$$

$$H = H_0\exp[i(\vec{q}\cdot\vec{r}-\omega t)] \tag{S7}$$

$$\Delta r_\alpha(jl,t) = \frac{e_\alpha(j)}{\sqrt{m_j}}\exp[i(\vec{q}\cdot\vec{r}(jl)-\omega t)] \tag{S8}$$

Substitute Eq. (S5-S8) into Eq. (S1-S4, 3, 4), we get

$$\vec{q}\times\vec{E}_0 = \mu_0\omega\vec{H}_0 \tag{S9}$$

$$\vec{q}\times\vec{H}_0 = -\omega\left(\varepsilon_0\vec{E}_0+\vec{P}_0\right) \tag{S10}$$

$$\vec{q}\cdot\left(\varepsilon_0\vec{E}_0+\vec{P}_0\right) = 0 \tag{S11}$$

$$\vec{q}\cdot\vec{H}_0 = 0 \tag{S12}$$

$$-\omega^2\sqrt{m_{j'}}e_\beta(j') = -\sum_{\alpha jl}\Phi_{\alpha\beta}(jl,j'l')*\frac{e_\alpha(j)}{\sqrt{m_j}}\exp[i\vec{q}\cdot\vec{r}(jl)-i\vec{q}\cdot\vec{r}(j'l')]+e\sum_{\gamma}Z_{\gamma\beta}(j'l')*E_{\gamma 0} \tag{S13}$$

$$P_{\beta 0} = \frac{e}{\Omega}\sum_{\mu j' l'} Z_{\beta\mu}(j'0)\frac{e_\mu(j')}{\sqrt{m_{j'}}}\exp[i\vec{q}\cdot\vec{r}(j'0) - i\vec{q}\cdot\vec{r}] + \sum_{\gamma}[\varepsilon_{\infty\beta\gamma} - \delta_{\beta\gamma}]\varepsilon_0 E_{\gamma 0} \quad (S14)$$

Eq. (S13) can be written as

$$\omega^2 e_\alpha(j) = \sum_{\beta j'} \mathrm{D}_{\alpha\beta}(jj', q)e_\beta(j') - \frac{e}{\sqrt{m_j}}\sum_{\gamma} Z_{\gamma\alpha}(j0) * E_{\gamma 0} \quad (S15)$$

where

$$\mathrm{D}_{\alpha\beta}(jj', \vec{q}) = \frac{1}{\sqrt{m_j m_{j'}}}\sum_{l'} \Phi_{\alpha\beta}(j0, j'l')\exp[i\vec{q}\cdot\vec{r}(j'l') - i\vec{q}\cdot\vec{r}(j0)] \quad (S16)$$

Eq. (S15) can be further written into matrix form

$$\left(\overleftrightarrow{D} - \omega^2\overleftrightarrow{I}\right)\vec{e} = e\overleftrightarrow{(Z_m)}^T\vec{E}_0 \quad (S17)$$

where $\vec{e}$ is a $3N \times 1$ matrix, $N$ is the number of atoms in a unit cell, $\overleftrightarrow{D}$ is a $3N \times 3N$ matrix depending on $\vec{q}$, $\overleftrightarrow{Z_m}$ is a $3 \times 3N$ matrix, whose element

$$(Z_m)_{\alpha,3(j-1)+\gamma} = \frac{Z_{\gamma\alpha}(j0)}{\sqrt{m_j}} \quad (S18)$$

When the wave length is much larger than the lattice constants (q ~ 0), Eq. (S14) can by simplified to

$$P_{\beta 0} = \frac{e}{\Omega}\sum_{\mu j'} Z_{\beta\mu}^{(j')}\frac{e_\mu(j')}{\sqrt{m_{j'}}} + \sum_{\gamma}[\varepsilon_{\infty\beta\gamma} - \delta_{\beta\gamma}]\varepsilon_0 E_{\gamma 0} \quad (S19)$$

Eq. (S19) can be further written into matrix form

$$\vec{P}_0 = \frac{e}{\Omega}\overleftrightarrow{(Z_m)}\ \vec{e} + \varepsilon_0\left(\overleftrightarrow{\varepsilon} - \overleftrightarrow{I}\right)\vec{E}_0 \quad (S20)$$

Substitute Eq. (S9) and Eq. (S20) into Eq. (S10), we get

$$\vec{q}\times\left(\vec{q}\times\vec{E}_0\right) = -\mu_0\omega^2\left[\frac{e}{\Omega}\overleftrightarrow{(Z_m)}\ \vec{e} + \varepsilon_0\overleftrightarrow{\varepsilon}\vec{E}_0\right] \quad (S21)$$

As

$$\vec{q}\times\left(\vec{q}\times\vec{E}_0\right) = \begin{pmatrix} q_x q_y E_y - q_y^2 E_x + q_x q_z E_z - q_z^2 E_x \\ -q_x^2 E_y + q_x q_y E_x + q_y q_z E_z - q_z^2 E_y \\ -q_x^2 E_z - q_y^2 E_z + q_x q_z E_x + q_y q_z E_y \end{pmatrix} = \begin{pmatrix} -q_y^2 - q_z^2 & q_x q_y & q_x q_z \\ q_x q_y & -q_x^2 - q_z^2 & q_y q_z \\ q_x q_z & q_y q_z & -q_x^2 - q_y^2 \end{pmatrix}\begin{pmatrix} E_x \\ E_y \\ E_z \end{pmatrix}, \quad (S22)$$

denote

$$\overleftrightarrow{F} = \begin{pmatrix} -q_y^2 - q_z^2 & q_x q_y & q_x q_z \\ q_x q_y & -q_x^2 - q_z^2 & q_y q_z \\ q_x q_z & q_y q_z & -q_x^2 - q_y^2 \end{pmatrix}, \quad (S23)$$

then Eq. (S21) can be written as

$$\left(\varepsilon_0\mu_0\omega^2\overleftrightarrow{\varepsilon} + \overleftrightarrow{F}\right)\vec{E}_0 + \mu_0\omega^2\frac{e}{\Omega}\overleftrightarrow{(Z_m)}\ \vec{e} = 0 \quad (S24)$$

Eq. (S24) and Eq. (S17) together is

$$\begin{pmatrix} \overleftrightarrow{D} - \omega^2\overleftrightarrow{I} & -e\overleftrightarrow{(Z_m)}^T \\ \mu_0\omega^2\frac{e}{\Omega}\overleftrightarrow{(Z_m)} & \varepsilon_0\mu_0\omega^2\overleftrightarrow{\varepsilon} + \overleftrightarrow{F} \end{pmatrix}\begin{pmatrix} \vec{e} \\ \vec{E}_0 \end{pmatrix} = 0 \quad (S25)$$

The condition for a nonzero solution of Eq. (S25) is that

$$\det\begin{pmatrix} \overleftrightarrow{D} - \omega^2\overleftrightarrow{I} & -e\overleftrightarrow{(Z_m)}^T \\ \mu_0\omega^2\frac{e}{\Omega}\overleftrightarrow{(Z_m)} & \varepsilon_0\mu_0\omega^2\overleftrightarrow{\varepsilon} + \overleftrightarrow{F} \end{pmatrix} = 0 \quad (S26)$$

This gives us $\omega$. Then we substitute $\omega$ into Eq. (S25) and get eigenvectors $\begin{pmatrix} \vec{e} \\ \vec{E}_0 \end{pmatrix}$.

**Section III.**

**Solutions of the generalized Huang's equation for 2D slab.**

The system is treated as a multilayer structure composed of three layers: substrate layer (z < -d), media layer (-d < z < 0), and air layer (z>0). The solution in the z > 0 air layers can be written as

$$E = E_{u0}\exp\left[i\left(\vec{k}_u \cdot \vec{r} - \omega t\right)\right] \quad (S27)$$

$$H = H_{u0}\exp\left[i\left(\vec{k}_u \cdot \vec{r} - \omega t\right)\right] \quad (S28)$$

The solution in the z < -d substrate layer can be written as

$$E = E_{d0}\exp\left[i\left(\vec{k}_d \cdot (\vec{r} + d\hat{z}) - \omega t\right)\right] \quad (S29)$$

$$H = H_{d0}\exp\left[i\left(\vec{k}_d \cdot (\vec{r} + d\hat{z}) - \omega t\right)\right] \quad (S30)$$

where $\vec{k}_u = \vec{q}_t + ik_{uz}\vec{z}$ and $\vec{k}_d = \vec{q}_t - ik_{dz}\vec{z}$; $\vec{q}_t = q_x\vec{i} + q_y\vec{j}$; $k_{uz} = \sqrt{q_x^2 + q_y^2 - \frac{\omega^2}{c^2}}$; $k_{dz} = \sqrt{q_x^2 + q_y^2 - \varepsilon_d\frac{\omega^2}{c^2}}$; $\varepsilon_d$ is the dielectric function of the substrate (usually $SiO_2$). Substutiting Eq. (S27, S28) into Eq. (S1-S4) yields only one independent equation in the air layer.

$$k_{ux}E_{u0x} + k_{uy}E_{u0y} + k_{uz}E_{u0z} = 0 \quad (S31)$$

Substituting Eq. (S29, S30) into Eq. (S1-S4) yields only one independent equation in the substrate layer.

$$k_{dx}E_{d0x} + k_{dy}E_{u0y} + k_{dz}E_{u0z} = 0 \quad (S32)$$

The solution in the media can be written as

$$\vec{P} = \vec{P}_{a0}\exp[i(\vec{q}_t \cdot \vec{r}_t + q_z z - \omega t)] + \vec{P}_{b0}\exp[i(\vec{q}_t \cdot \vec{r}_t - q_z z - \omega t)] \quad (S33)$$

$$\vec{E} = \vec{E}_{a0}\exp[i(\vec{q}_t \cdot \vec{r}_t + q_z z - \omega t)] + \vec{E}_{b0}\exp[i(\vec{q}_t \cdot \vec{r}_t - q_z z - \omega t)] \quad (S34)$$

$$\vec{H} = \vec{H}_{a0}\exp[i(\vec{q}_t \cdot \vec{r}_t + q_z z - \omega t)] + \vec{H}_{b0}\exp[i(\vec{q}_t \cdot \vec{r}_t - q_z z - \omega t)] \quad (S35)$$

where $\vec{r}_t = x\vec{i} + y\vec{j}$. The *a* component and the *b* component propagate along the *z* and -*z* directions, respectively. The atomic vibration also has two component

$$\Delta r_\alpha(jl, t) = \frac{e_\alpha(j)}{\sqrt{m_j}}\exp[i(\vec{q}_t \cdot \vec{r}_t + q_z z - \omega t)] \quad (S36)$$

$$\Delta r_\beta(jl, t) = \frac{g_\beta(j)}{\sqrt{m_j}}\exp[i(\vec{q}_t \cdot \vec{r}_t - q_z z - \omega t)] \quad (S37)$$

Let $\vec{q}_1 = \vec{q}_t + q_z\hat{z}$ and $\vec{q}_2 = \vec{q}_t - q_z\hat{z}$ and substitute the *a* and *b* components of Eq. (S33-S37) into Eq. (S1-S4) and Eq. (3-4), we get

$$\vec{q}_1 \times \vec{E}_{a0} = \mu_0\omega\vec{H}_{a0} \quad (S38)$$

$$\vec{q}_2 \times \vec{E}_{b0} = \mu_0\omega\vec{H}_{b0} \quad (S39)$$

$$\vec{q}_1 \times \vec{H}_{a0} = -\omega\left(\varepsilon_0\vec{E}_{a0} + \vec{P}_{a0}\right) \quad (S40)$$

$$\vec{q}_2 \times \vec{H}_{b0} = -\omega\left(\varepsilon_0\vec{E}_{b0} + \vec{P}_{b0}\right) \quad (S41)$$

$$\vec{q}_1 \cdot \left(\varepsilon_0\vec{E}_{a0} + \vec{P}_{a0}\right) = 0 \quad (S42)$$

$$\vec{q}_2 \cdot \left(\varepsilon_0\vec{E}_{b0} + \vec{P}_{b0}\right) = 0 \quad (S43)$$

$$\vec{q}_1 \cdot \vec{H}_{a0} = 0 \quad (S44)$$

$$\vec{q}_2 \cdot \vec{H}_{b0} = 0 \quad (S45)$$

$$-\omega^2\sqrt{m_{j'}}e_\delta(j')\exp[i(\vec{q}_t \cdot \vec{r}_t(j'l') + q_z z(j'l') - \omega t)]$$

$$= -\sum_{\alpha jl}\Phi_{\alpha\delta}(jl, j'l') * \frac{e_\alpha(j)}{\sqrt{m_j}}\exp[i(\vec{q}_t \cdot \vec{r}_t(jl) + q_z z(jl) - \omega t)]$$

$$+ e\sum_\gamma Z_{\gamma\delta}(j'l') * E_{a0\gamma}\exp[i(\vec{q}_t \cdot \vec{r}_t(j'l') + q_z z(j'l') - \omega t)] \quad (S46)$$

$$-\omega^2\sqrt{m_{j'}}g_\sigma(j')\exp[i(\vec{q}_t\cdot\vec{r}_t(j'l')-q_z z(j'l')-\omega t)]$$

$$=-\sum_{\alpha jl}\Phi_{\beta\sigma}(jl,j'l')*\frac{g_\beta(j)}{\sqrt{m_j}}\exp[i(\vec{q}_t\cdot\vec{r}_t(jl)-q_z z(jl)-\omega t)]$$

$$+e\sum_{\gamma}Z_{\gamma\sigma}(j'l')*E_{b0\gamma}\exp[i(\vec{q}_t\cdot\vec{r}_t(j'l')-q_z z(j'l')-\omega t)] \quad (S47)$$

$$P_{a0\delta}\exp[i(\vec{q}_t\cdot\vec{r}_t(j'l')+q_z z(j'l')-\omega t)]$$

$$=\frac{e}{\Omega}\sum_{\mu j'l}Z_{\delta\mu}^{(j')}\frac{e_\mu(j')}{\sqrt{m_{j'}}}\exp[i(\vec{q}_t\cdot\vec{r}_t(j'l')+q_z z(j'l')-\omega t)]$$

$$+\sum_{\gamma}[\varepsilon_{\infty\delta\gamma}-\delta_{\delta\gamma}]\varepsilon_0 E_{a0\gamma}\exp[i(\vec{q}_t\cdot\vec{r}_t(j'l')+q_z z(j'l')-\omega t)] \quad (S48)$$

$$P_{b0\sigma}(z)=\frac{e}{\Omega}\sum_{\mu j'l'}Z_{\sigma\mu}^{(j')}\frac{g_\mu(j')}{\sqrt{m_{j'}}}\exp[i(\vec{q}_t\cdot\vec{r}_t(j'l')-q_z z(j'l')-\omega t)]$$

$$+\sum_{\gamma}[\varepsilon_{\infty\sigma}\ -\delta_{\sigma\gamma}]\varepsilon_0 E_{b0\gamma}\exp[i(\vec{q}_t\cdot\vec{r}_t(j'l')-q_z z(j'l')-\omega t)] \quad (S49)$$

Substitute Eq. (S40) into Eq. (S38), and we get

$$q_x q_y E_{a0y}-q_y^2E_{a0x}+q_x q_z E_{a0z}-q_z^2E_{a0x}=-\mu_0\varepsilon_0\omega^2E_{a0x}-\mu_0\omega^2P_{a0x} \quad (S50)$$

$$-q_x^2E_{a0z}-q_y^2E_{a0z}+q_x q_z E_{a0x}+q_y q_z E_{a0y}=-\mu_0\varepsilon_0\omega^2E_{a0z}-\mu_0\omega^2P_{a0z} \quad (S51)$$

Substitute Eq. (S41) into Eq. (S39), and we get

$$q_x q_y E_{b0y}-q_y^2E_{b0x}-q_x q_z E_{b0z}-q_z^2E_{b0x}=-\mu_0\varepsilon_0\omega^2E_{b0x}-\mu_0\omega^2P_{b0x} \quad (S52)$$

$$-q_x^2E_{b0z}-q_y^2E_{b0z}-q_x q_z E_{b0x}-q_y q_z E_{b0y}=-\mu_0\varepsilon_0\omega^2E_{b0z}-\mu_0\omega^2P_{b0z} \quad (S53)$$

As there is only p-polarization (electric field polarition direction parallel to the plane of incidence) in the s-SNOM experiment, we choose the coordinate system that the $\vec{q}$ and $\vec{E}$ are in the *x-z* plane, and the coordinate system is the principal axes system. The continuity of the electric field in the *x* direction at $z = 0$ and $z = -d$ lead to

$$E_{a0x}+E_{b0x}=E_{u0x} \quad (S54)$$

$$E_{a0x}\exp(-iq_z d)+E_{b0x}\exp(iq_z d)=E_{d0x} \quad (S55)$$

The continuity of magnet intensity in the *y* direction at $z = 0$ and $z = -d$ leads to

$$k_{uz}E_{u0x}-q_x E_{u0z}=q_z E_{a0x}-q_x E_{a0z}-q_z E_{b0x}-q_x E_{b0z} \quad (S56)$$

$$k_{dz}E_{d0x}-q_x E_{d0z}=(q_z E_{a0x}-q_x E_{a0z})\exp(-iq_z d)+(-q_z E_{b0x}-q_x E_{b0z})\exp(iq_z d) \quad (S57)$$

Eq. (S46) can be simplified to

$$\omega^2 e_\alpha(j)=\sum_{\delta j'}\mathrm{D}_{\alpha\delta}(jj',\vec{q}_1)e_\delta(j')-e\sum_{\gamma}\frac{Z_{\gamma\alpha}(j0)}{\sqrt{m_j}}*E_{a0\gamma} \quad (S58)$$

Write it in matrix form

$$\left(\overleftrightarrow{D}(\vec{q}_1)-\omega^2\overleftrightarrow{I}_{3N}\right)\vec{e}=e\overleftrightarrow{Z_m}^{T}\vec{E}_{a0} \quad (S59)$$

where $\overleftrightarrow{I}_{3N}$ is $3N\times 3N$ identity matrix. Similarly, Eq. (S47) can be written in matrix form

$$\left(\overleftrightarrow{D}(\vec{q}_2)-\omega^2\overleftrightarrow{I}_{3N}\right)\vec{g}=e\overleftrightarrow{Z_m}^{T}\vec{E}_{b0} \quad (S60)$$

where $\vec{g}$ is a $3N\times 1$ matrix.

When q ~ 0, Eq. (S48) can be simplified to

$$P_{a0\delta}=\frac{e}{\Omega}\sum_{\mu j'l'}Z_{\delta\mu}^{(j')}\frac{e_\mu(j')}{\sqrt{m_{j'}}}+\sum_{\gamma}[\varepsilon_{\infty\delta\gamma}-\delta_{\delta\gamma}]\varepsilon_0 E_{a0\gamma} \quad (S61)$$

Write it in matrix form

$$\vec{P}_{a0} = \frac{e}{\Omega}\overleftrightarrow{Z_m}\vec{e} + \varepsilon_0\left(\overleftrightarrow{\varepsilon} - \overleftrightarrow{I_3}\right)\vec{E}_{a0} \tag{S62}$$

where $\overleftrightarrow{\varepsilon}$ is a $3N \times 3N$ matrix, $\overleftrightarrow{I_3}$ is a $3 \times 3$ identity matrix. Similarly, Eq. (S49) can be written in matrix form

$$\vec{P}_{b0} = \frac{e}{\Omega}\overleftrightarrow{Z_m}\vec{g} + \varepsilon_0\left(\overleftrightarrow{\varepsilon} - \overleftrightarrow{I_3}\right)\vec{E}_{b0} \tag{S63}$$

Eqs. (S31, S32, S50-S57, S59, S60, S62, S63) are a system of linear homogeneous equations. The unknowns are $\vec{e}$, $\vec{g}$, $\vec{E}_{a0}$, $\vec{E}_{b0}$, $\vec{P}_{a0}$, $\vec{P}_{b0}$, $\vec{E}_{u0}$ and $\vec{E}_{d0}$, in which there are $6N$+12 unkown components (the $\vec{E}$ and $\vec{P}$ components in the $y$ direction are fixed to zero). The coefficient matrix the linear equations is

$$\begin{pmatrix}
\overleftrightarrow{D}(\vec{q}_1) - \omega^2\overleftrightarrow{I}_{3N} & & -e\overleftrightarrow{Z_m}^T & & & & & \\
& \overleftrightarrow{D}(\vec{q}_2) - \omega^2\overleftrightarrow{I}_{3N} & & -e\overleftrightarrow{Z_m}^T & & & & \\
\frac{e}{\Omega}\overleftrightarrow{Z_m} & & \varepsilon_0\left(\overleftrightarrow{\varepsilon} - \overleftrightarrow{I_3}\right) & & -\overleftrightarrow{I_3} & & & \\
& \frac{e}{\Omega}\overleftrightarrow{Z_m} & & \varepsilon_0\left(\overleftrightarrow{\varepsilon} - \overleftrightarrow{I_3}\right) & & -\overleftrightarrow{I_3} & & \\
& & \begin{matrix} \mu_0\varepsilon_0\omega^2 - q_z^2 & 0 & q_xq_z \\ q_xq_z & 0 & \mu_0\varepsilon_0\omega^2 - q_x^2 \end{matrix} & & \begin{matrix} \mu_0\omega^2 & 0 & 0 \\ 0 & 0 & \mu_0\omega^2 \end{matrix} & & & \\
& & & \begin{matrix} \mu_0\varepsilon_0\omega^2 - q_z^2 & 0 & -q_xq_z \\ -q_xq_z & 0 & \mu_0\varepsilon_0\omega^2 - q_x^2 \end{matrix} & & \begin{matrix} \mu_0\omega^2 & 0 & 0 \\ 0 & 0 & \mu_0\omega^2 \end{matrix} & & \\
& & & & & & \begin{matrix} q_x & 0 & k_{uz} \end{matrix} & \\
& & & & & & & \begin{matrix} q_x & 0 & k_{dz} \end{matrix} \\
& & \begin{matrix} 1 & 0 & 0 \\ e^{-iq_zd} & 0 & 0 \\ q_z & 0 & -q_x \\ q_ze^{-iq_zd} & 0 & -q_xe^{-iq_zd} \end{matrix} & \begin{matrix} 1 & 0 & 0 \\ e^{iq_zd} & 0 & 0 \\ -q_z & 0 & -q_x \\ -q_ze^{iq_zd} & 0 & -q_xe^{iq_zd} \end{matrix} & & & \begin{matrix} -1 & 0 & 0 \\ & & \\ -k_{uz} & 0 & q_x \\ & & \end{matrix} & \begin{matrix} & & \\ -1 & 0 & 0 \\ & & \\ -k_{dz} & 0 & q_x \end{matrix}
\end{pmatrix}$$

The condition for a nonzero solution is that the rank of the coefficient matrix is less than $6N$+12. This gives us $\omega$ and $q_z$ depending on $q_x$. Then we substitute $\omega$ and $q_z$ into the system of linear homogeneous equations and get eigenvectors $\vec{e}$, $\vec{g}$, $\vec{E}_{a0}$, $\vec{E}_{b0}$, $\vec{P}_{a0}$, $\vec{P}_{b0}$, $\vec{E}_{u0}$ and $\vec{E}_{d0}$.

**Section. IV.**

We can calculate the circular polarization of the ion vibration component of these eigenstates by representing the ion vibration component of the eigenstate [6, 7]

$$\epsilon = (x_1\, y_1 \cdots x_N\, y_N)^T, \tag{S64}$$

where $N$ is the number of atoms in a unit cell. And by defining a set of right/left circularly polarized basis

$$|R_1\rangle \equiv \frac{1}{\sqrt{2}}(1 \;\; i \;\; 0 \;\; 0 \;\cdots)^T, |L_1\rangle \equiv \frac{1}{\sqrt{2}}(1 \;\; -i \;\; 0 \;\; 0 \;\cdots)^T,$$

$$|R_\alpha\rangle \equiv \frac{1}{\sqrt{2}}(0 \;\; 0 \;\cdots 1 \;\; i \;\cdots 0 \;\; 0)^T, |L_\alpha\rangle \equiv \frac{1}{\sqrt{2}}(0 \;\; 0 \;\cdots 1 \;\; -i \;\cdots 0 \;\; 0)^T,$$

$$|R_N\rangle \equiv \frac{1}{\sqrt{2}}(0 \;\; 0 \;\cdots 1 \;\; i)^T, |L_N\rangle \equiv \frac{1}{\sqrt{2}}(0 \;\; 0 \;\cdots 1 \;\; -i)^T,$$

we can decompose each ion vibration component of the eigenstate into right and left circularly polarized components:

$$\epsilon = \sum_{\alpha=1}^{N} \epsilon_{R_\alpha}|R_\alpha\rangle + \epsilon_{L_\alpha}|L_\alpha\rangle. \tag{S65}$$

Then the circular polarization along $z$ is

$$s_z = \hbar\sum_{\alpha=1}^{N}\left(\left|\epsilon_{R_\alpha}\right|^2 + \left|\epsilon_{L_\alpha}\right|^2\right). \tag{S66}$$

**Section V.**

**Growth of the α-$MoO_3$ 2D Flakes:** The α-$MoO_3$ 2D flakes were synthesized by thermal physical deposition method[8]. Briefly, an alumina crucible with $MoO_3$ powder (80 mg) as the source was placed at the center of a quartz tube. The $SiO_2$ substrate was cleaned and placed at the low-temperature region of 650 °C. The separation between the source and substrate was 13.0−14.5 cm. The source was heated up to 780 °C in 40 min and then kept at that temperature for another 10 min. The substrate was heated up to 650 °C in 35 min and then kept at that temperature for another 15 min. During the thermal treatment, the $MoO_3$ powder was sublimated and recrystallized onto the Si substrate. Subsequently, the quartz tube was cooled down naturally to room temperature, whereby numerous α-$MoO_3$ 2D flakes with different thicknesses (from 70 nm to a few hundred nm) were found on the $SiO_2$ substrate. We reduce the thickness to 22 nm with reactive ion etching (RIE). The flow of $CHF_3$, $O_2$ and Ar is 30 sccm, 6 sccm and 15 sccm, respectively.

**The s-SNOM nano-imaging** was conducted using a scattering-type near-field optical microscope (NeaSNOM, Neaspec GmbH). To image the PhPs in real space, a mid-infrared laser (Access Laser) with tunable wavelengths from 9.20 to 10.99 µm (910–1087 cm–1) was focused onto the sample with a metal-coated AFM tip (Arrow-IrPt, Nanoworld). During the measurements, the tip was vibrated vertically with a frequency of around 280 kHz. The back-scattered light from the tip was demodulated and detected at a fourth harmonic of the tip vibration frequency.

**Section. VI.**

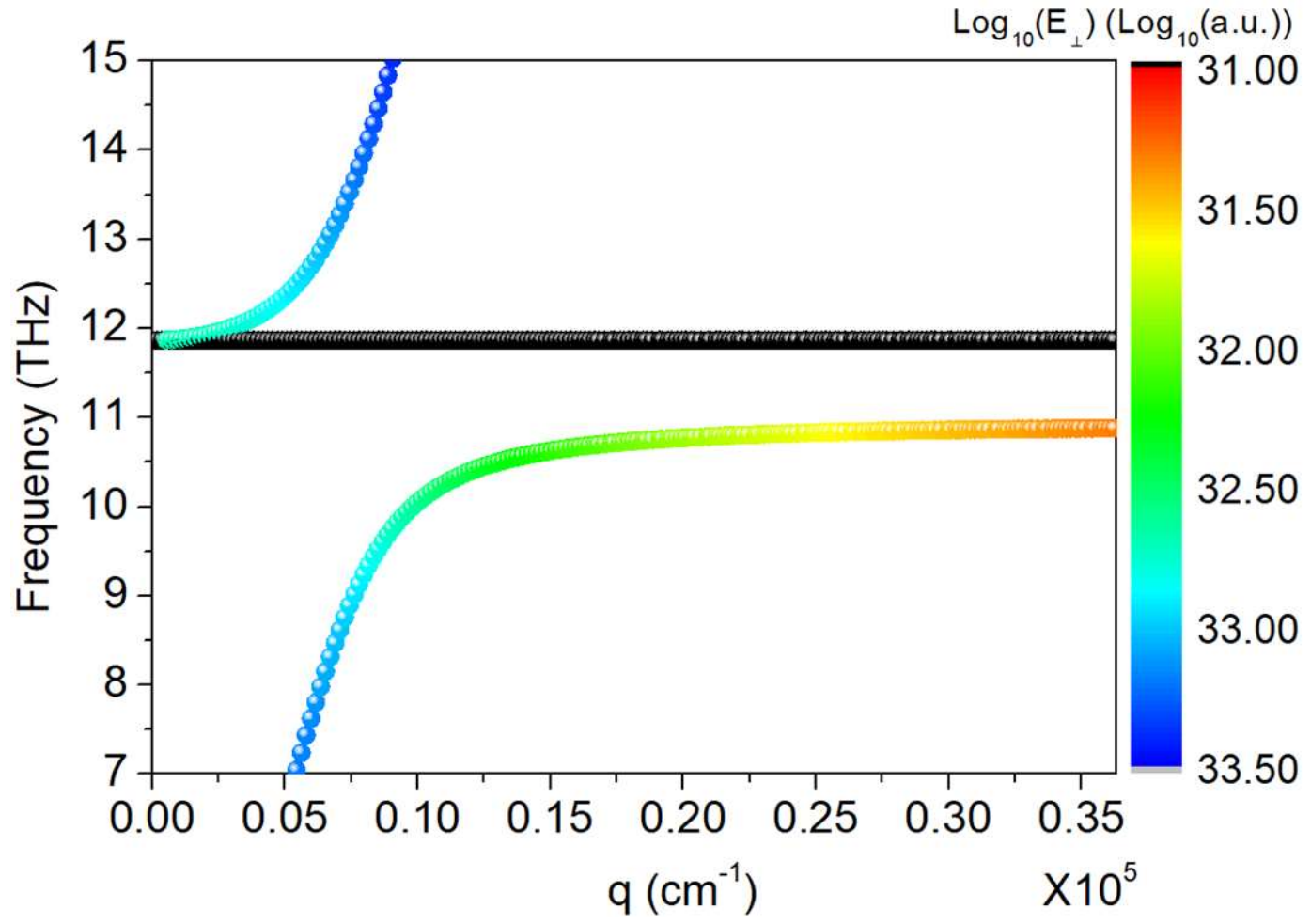

Figure S1. PhP dispersion in GaP. The color indicates the electric field intensity perpendicular to the wave vector direction.

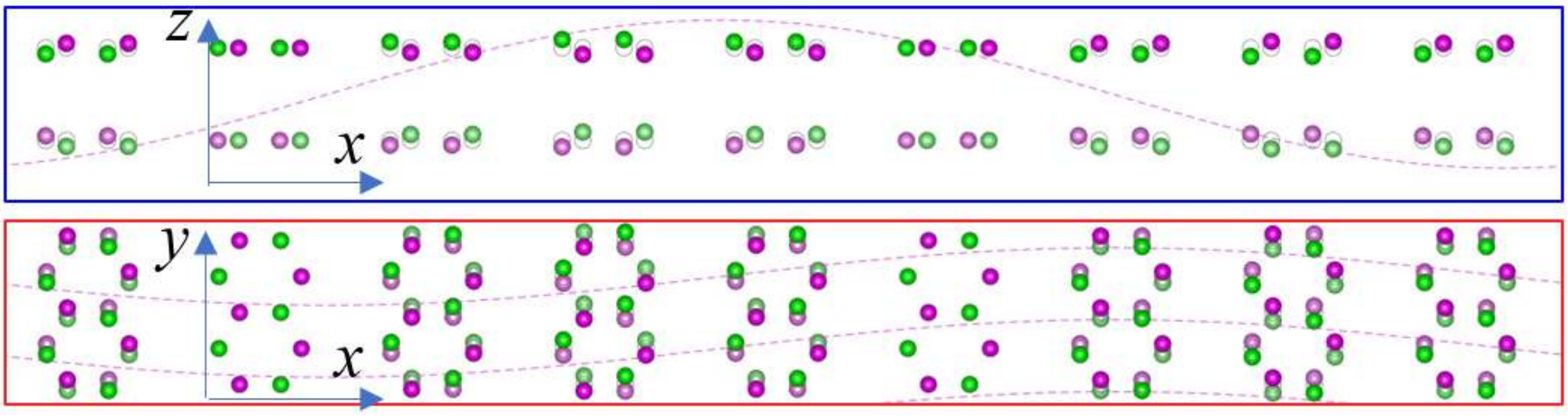

Figure S2. Atomic vibration in bulk hBN. The top panel corresponds to the blue symbols in Figure 1b. The bottom panel corresponds to the red symbols in Figure 1c. The purple (green) balls refer to boron (nitrogen) atoms. The wave vector is in the *x*-direction. Because the wavelength of PhP is much longer than the lattice constants, the ions along the x-direction are depicted only in every *N* lattice, with *N* being of the order of a thousand or more.

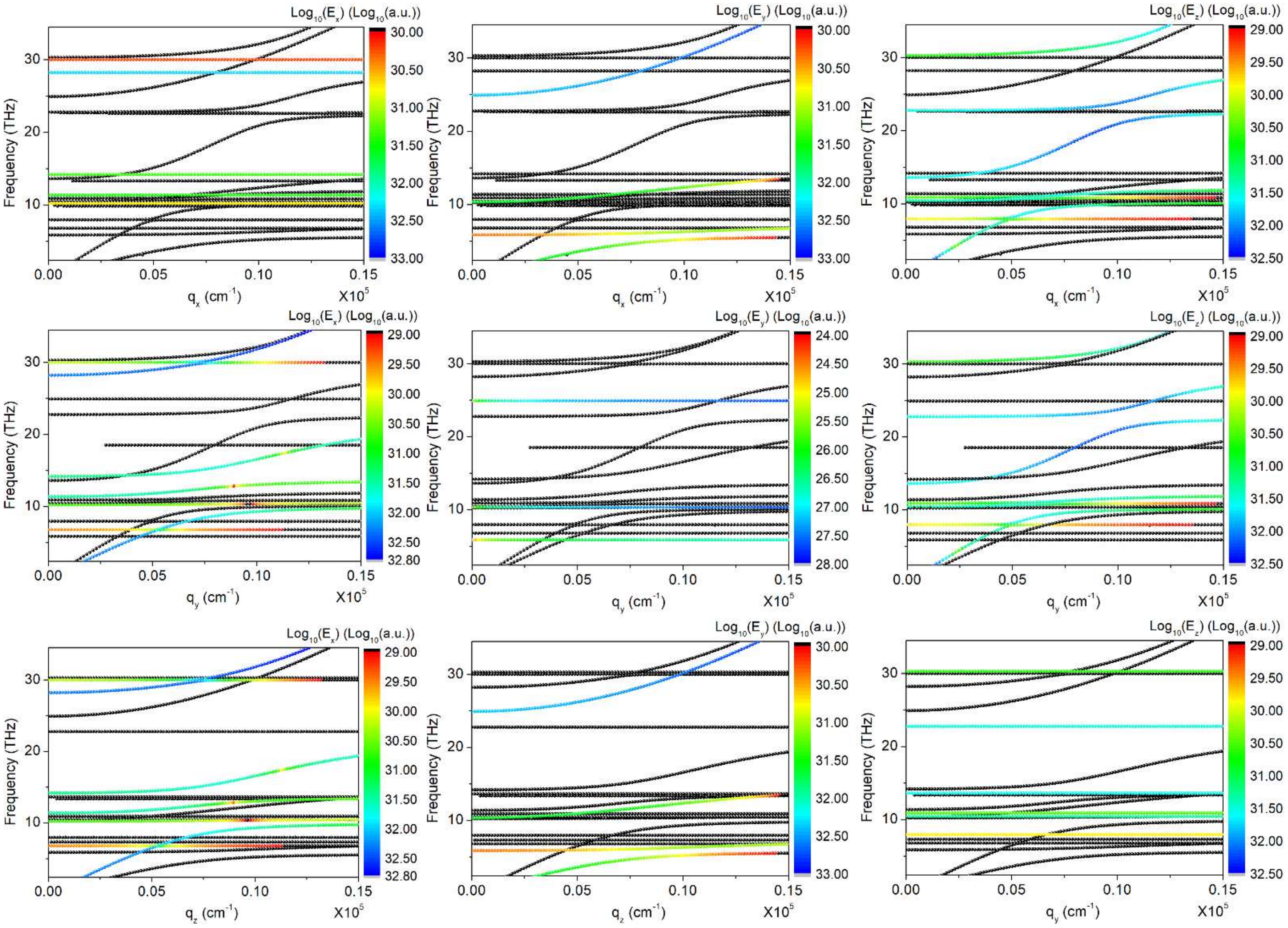

Figure S3. Electric field intensity of PhPs in bulk α-MoO3 when the wave vector is along the *x*-direction (top panels), *y*-direction (middle panels) and *z*-direction (bottom panels). The color of the symbols indicates the electric field intensity in the *x*-direction (left panels), *y*-direction (middle panels) and *z*-direction (right panels).

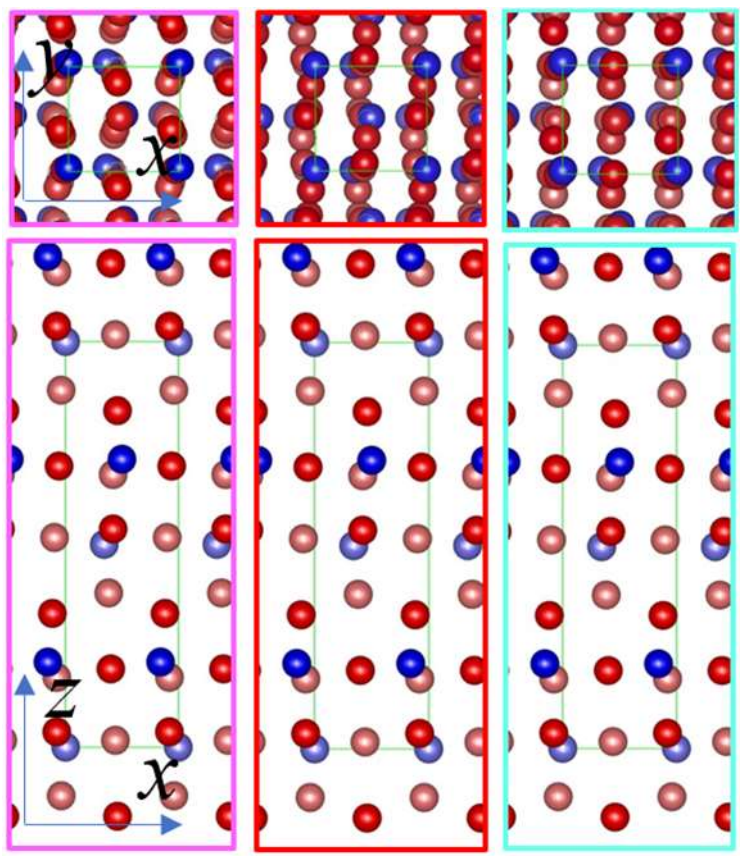

Figure S4. The *y*-direction atomic vibration of PhPs whose wave vector is along [100] (denoted as *x*) or [010] (denoted as *z*) direction in bulk α-$MoO_3$. The left, middle and right panels correspond to the pink, red and cyan symbols in Figure 1d and 1f, respectively. The top (bottom) panels are viewed from the *z* (*y*) direction. The green rectangles refer to the unit cell.

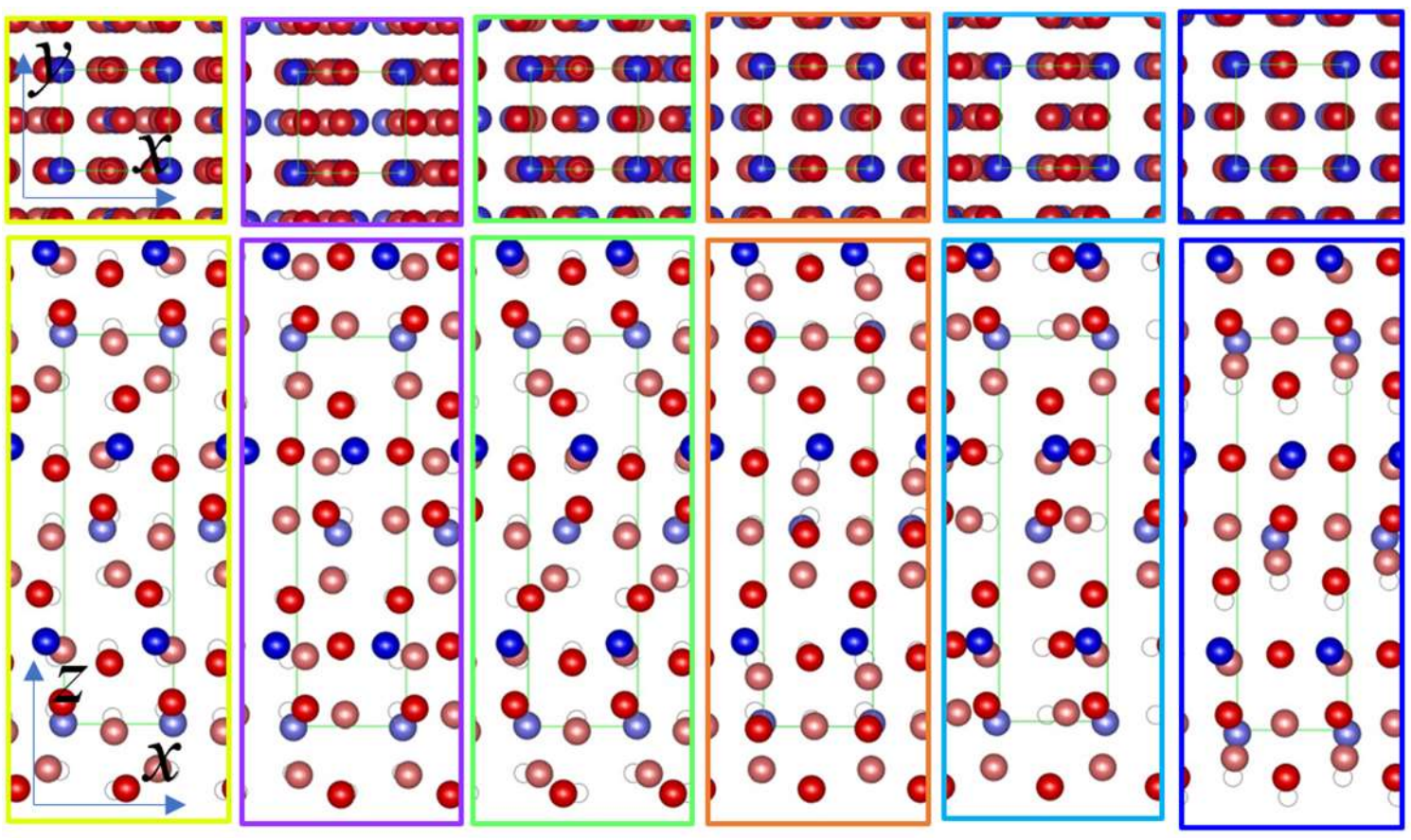

Figure S5. The *xz* plane atomic vibration of PhPs whose wave vector is along [100] (denoted as *x*) direction in bulk α-$MoO_3$. The colors of the rectangles surrounding the panels correspond to the colors of the symbols in Figure 1d. The top (bottom) panels are viewed from the *z* (*y*) direction. The green rectangles refer to the unit cell.

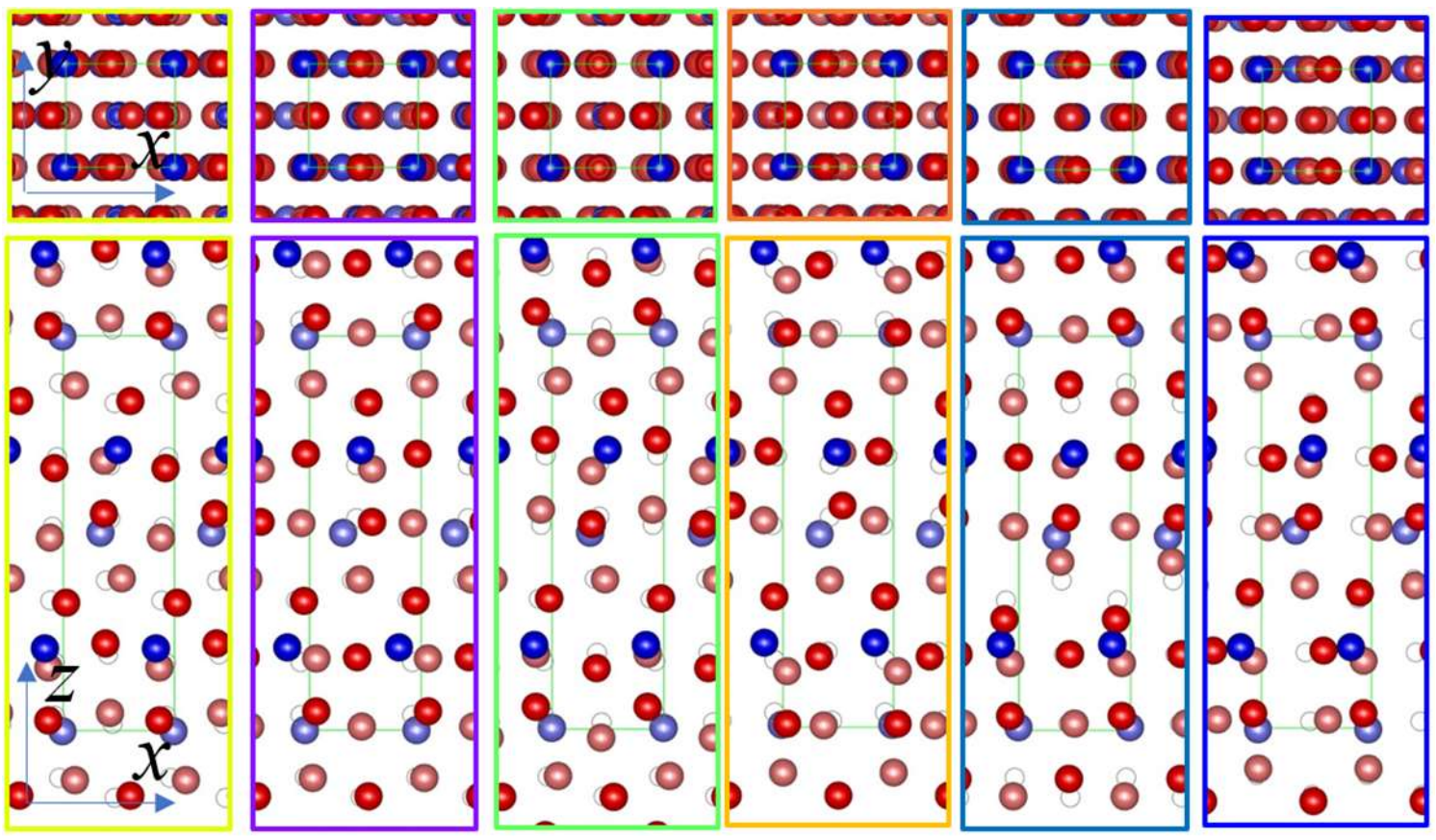

Figure S6. The *xz* plane atomic vibration of PhPs whose wave vector is along [010] (denoted as *z*) direction in bulk α-$MoO_3$. The colors of the rectangles surrounding the panels correspond to the colors of the symbols in Figure 1f. The top (bottom) panels are viewed from the *z* (*y*) direction. The green rectangles refer to the unit cell.

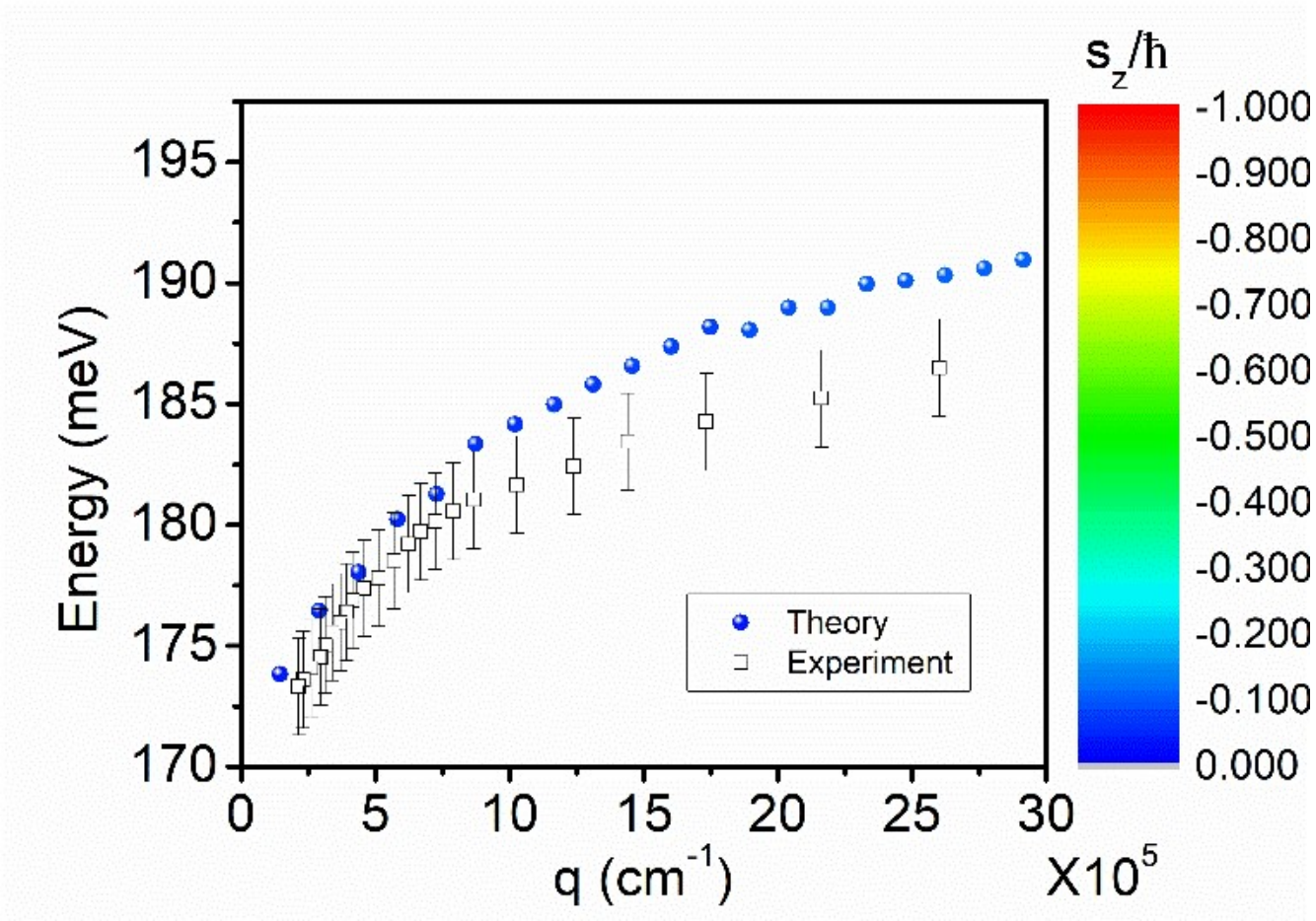

Figure S7. Experimental (black empty squares) and theoretical (color balls) PhP dispersion in hBN flake with thickness 3 nm. The color indicates the circular polarizations of the ion vibration component of the PhP eigenstates. The experimental data are from Ref. [9]